\documentclass{article}
\usepackage{spconf}
\usepackage{mypre}
\usepackage{mytikzlib}
\usepackage[bottom]{footmisc}

\title{Spatial Attention for Far-field Speech Recognition\\with Deep Beamforming Neural Networks}
\name{%
  Weipeng He$^{1,2}$, Lu Lu$^{3}$, Biqiao Zhang$^{3}$, Jay Mahadeokar$^{3}$, Kaustubh Kalgaonkar$^{3}$, Christian Fuegen$^{3}$
  \thanks{*This work was done when the first author was an intern with Facebook.}
}
\address{%
  $^{1}$Idiap Research Institute, Switzerland \\
  $^{2}$\'Ecole Polytechnique F\'ed\'erale de Lausanne (EPFL), Switzerland \\
  $^{3}$Facebook, 1 Hacker Way, Menlo Park, CA 94025, USA \\
  {
    \small
    \texttt{weipeng.he@idiap.ch}\hspace{4mm}%
    \texttt{lulu0314@gmail.com}\hspace{4mm}%
    \texttt{\{didizbq,jaym,kaustubhk,fuegen\}@fb.com}
  }
}

\begin{document}
\ninept%
\maketitle
\begin{abstract}
  In this paper, we introduce spatial attention for refining the information
  in multi-direction neural beamformer for far-field automatic speech recognition.
  Previous approaches of neural beamformers with multiple look directions,
  such as the factored complex linear projection, have shown promising results.
  However, the features extracted by such methods contain redundant information,
  as only the direction of the target speech is relevant.
  We propose using a spatial attention subnet to weigh the features from different directions,
  so that the subsequent acoustic model could focus on the most relevant features for the speech recognition.
  Our experimental results show that spatial attention achieves up to 9\% relative word error rate improvement over methods without the attention.
\end{abstract}
\begin{keywords}
  Deep beamforming networks, multi-channel far-field speech recognition, array signal processing, attention
\end{keywords}

\section{Introduction}

Smart speakers have become increasingly popular in people's daily life,
and far-field Automatic Speech Recognition (ASR) is one of the important technologies behind them.
Far-field ASR is more challenging than near-field ASR as the far-field signal is more corrupted by reverberation, background noise, as well as interfering voices.
In this paper, we address the far-field ASR problem by introducing deep beamforming neural networks with spatial attention, which show improvements over existing methods.

Previous studies have applied traditional signal processing techniques to the corrupted signal before the ASR\@, such as noise reduction~\cite{boll_suppression_1979},
dereverberation~\cite{yoshioka_generalization_2012},
beamforming~\cite{cox_robust_1987,lacroix_superdirective_2001},
and post-filtering~\cite{lacroix_post-filtering_2001,mccowan_microphone_2003}.
Among them, beamforming exploits the spatial information in the multi-channel signal,
so that the sounds from different directions can be enhanced or suppressed.
Technically, the beamformers apply filtering on the input and summation across channels.
The filter coefficients (i.e., beamformer weights) are chosen
by solving objective functions that make the enhanced signal better quality under certain assumptions.
The performance of the traditional beamforming techniques is often limited in complex real-world environments. This is because that these methods often rely heavily on the assumptions of some ideal conditions about the environments,
such as stationary signal, high signal-to-noise ratio (SNR), and precisely estimated direction of arrival or steering vectors.
Furthermore, their objective functions are indirect with regard to ASR\@.
Therefore, the enhanced signal does not necessarily improve the ASR results.

More recently, deep neural networks for beamforming have been shown to outperform traditional methods,
as they do not require strong assumption about the environments.
These deep beamforming network approaches may be divided into three main categories:
\begin{enumerate}
\itemsep0em
  \item \emph{Mask estimation}~\cite{erdogan_multi-channel_2016,heymann_neural_2016,ochiai_unified_2017}:
    the neural network predicts the speech mask (and noise mask),
    which is then used for estimating the covariance matrix for Minimum Variance Distortionless Response (MVDR) or Generalized Eigenvector (GEV) beamforming.
  \item \emph{Adaptive filter coefficients estimation on single look direction}~\cite{xiao_deep_2016,li_neural_2016,meng_deep_2017,ochiai_unified_2017,qian_deep_2018}:
    the neural network predicts the beamforming weights as an intermediate output.
    The weights enhance the signal in one look direction.
  \item \emph{Fixed filter coefficients estimation for beamforming on multiple directions}~\cite{sainath_factored_2016,sainath_reducing_2016,li_acoustic_2017}:
    the beamforming weights are part of the neural network parameters.
    They are optimized to extract features on multiple directions and kept fixed after training.
\end{enumerate}

Among these approaches, the factored complex linear projection (fCLP) from the third category
has shown promising results with real-world smart speakers~\cite{li_acoustic_2017}.
This approach is advantageous as it is computational efficient,
does not require Direction of Arrival (DOA) estimation,
and jointly optimizes enhancement and recognition.
In fCLP, the features on all look directions are used as the input of the back-end network for Acoustic Modeling (AM).
However, such features contain redundant information,
as ideally only the features from the direction of the target speech are useful for the recognition.

In this paper, we propose using spatial attention to refine the multi-direction features in the fCLP approach.
The spatial attention, computed from multi-directional features, indicates how informative each direction is for recognizing the target speech.
We use attention score to weigh the features from the multiple directions using average pooling. This allows the subsequent acoustic model to focus on the features most relevant to ASR and reduces the number of parameters of the network, compared to the original multi-directional setting.
The experimental results show that there is a significant improvement by adding spatial attention to the original fCLP approach.

Previous studies have explored similar ideas by using attention to select the most informative input channels for multi-channel far-field ASR~\cite{kim_recurrent_2016,braun_multi-channel_2018}.
Our approach differs from them as we apply attention weighting to the spatial filters, not to the input channels.

\section{Approach}

\begin{figure}[t]
  \begin{center}
    \centering \sf \scriptsize \mathversion{sfnums}
    \includegraphics{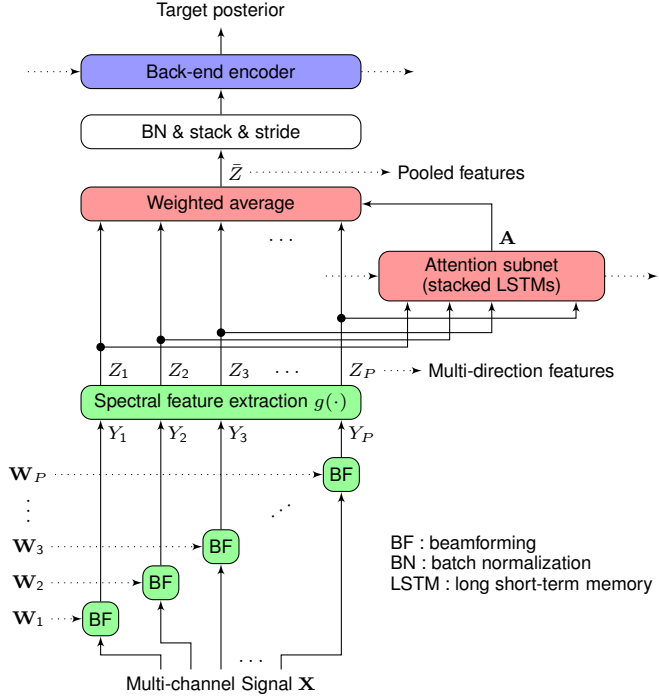}
  \end{center}
  \caption{%
    Deep beamforming neural network with spatial attention, consisting of the neural beamformer (green), the spatial attention subnet (red), and the back end (blue).
    This figure shows the processing of a single frame,
    and the time dependency is indicated by the dotted arrows.
  }\label{fig:net_arch}
\end{figure}

We propose an end-to-end neural network (Figure~\ref{fig:net_arch})
for acoustic modeling from raw multi-channel signals with three components:
\begin{itemize}\itemsep0em
  \item \emph{Neural beamformer}, which extracts speech features on multiple look directions.
  \item \emph{Attention-based pooling module}, which picks the most informative features by spatial attention.
  \item \emph{Back end}, which predicts the sequence of target posteriors
  (e.g., context-dependent phones or graphemes)
  from the sequence of pooled features.
\end{itemize}
The three components are trained jointly,
so that the speech enhancement and feature extraction front-end are directly optimized to reduce the target classification error.

\subsection{Front end: Neural Beamformer}

The neural beamformer front end is based on the factored Complex Linear Projection (fCLP)~\cite{sainath_reducing_2016,li_acoustic_2017}.
Specifically, the front-end takes multi-channel short-time Fourier transform (STFT)
$\mathbf{X}[t,f] = {\left(X_1[t,f],\ldots,X_M[t,f]\right)}^{\text{\sffamily T}} \in \mathbb{C}^M$
as the network input.
$t$ and $f$ are the indices of the time and frequency bin, respectively,
$M$ is the number of microphones, and $^{\text{\sffamily T}}$ is the matrix transpose.
The input signal is beamformed with $P$ spatial filters (look directions)
using the weights $\mathbf{W}_p[f] \in \mathbb{C}^M$:
\begin{equation}
  Y_p[t,f] = \mathbf{W}_p^{\text{\sffamily H}}[f] \mathbf{X}[t,f], \quad p = 1,2,\ldots,P,
  \label{eq:beamform}
\end{equation}
where $p$ is the direction index and ${}^{\text{\sffamily H}}$ is the conjugate transpose.
The beamforming weights $\mathbf{W}_p[f]$ are parameters of the neural network.

After the spatial filtering (beamforming), we extract $L$ spectral features from each enhanced signal $Y_p$:
\begin{equation}
  Z_{p,l}[t] = g_l(Y_p[t]), \quad l = 1,2,\dots,L,
  \label{eq:specfil}
\end{equation}
where $Y_p[t] = {(Y_p[t,1],\ldots,Y_p[t,F])}^{\text{\sffamily T}}$.
In our experiments, we use the Complex Linear Projection (CLP)~\cite{variani_complex_2016} as the spectral features extractor.
That is:
\begin{equation}
  Z_{p,l}[t] = \log \left| \sum_{f=1}^{F} Y_p[t,f] G_l[f] \right|,
  \label{eq:clp}
\end{equation}
where $G_l = {\left( G_l[1],\ldots,G_l[F] \right)}^{\text{\sffamily T}} \in \mathbb{C}^F$ are the parameters of the network.
We have tested another feature extraction method with multi-layer perceptron
in the preliminary experiments but have not found significant difference.
The output of the beamformer front-end on each frame
is a 2-D tensor with the dimensions of the directions and the spectral features,
that is $Z[t] = \left\{Z_{p,l}\right\} \in \mathbb{R}^{P \times L}$.

Note that while the functions are complex-valued,
the network computes the real and imaginary parts as separate real-valued functions,
which allow us to calculate the derivative of the loss for back-propagation.

\begin{figure}[t]
  \begin{center}
    \sf \scriptsize \mathversion{sfnums}
    \includegraphics{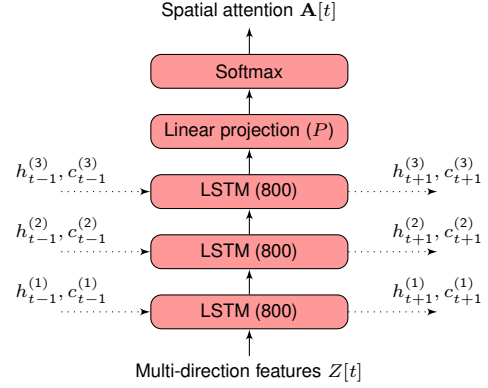}
  \end{center}
  \caption{%
    The attention subnet.
    $h$ and $c$ are the hidden and cell states of the LSTMs.
    For online attention, the output $\mathbf{A}[t]$ is smoothed by averaging over time (omitted in the figure).
    For offline attention, the output is taken at the last frame.
  }\label{fig:online_att}
\end{figure}

\subsection{Spatial Attention}

We use an attention-based pooling module to select the most informative direction.
Given that the target speech is from a specific direction,
only one or a few out of the $P$ directions are relevant for recognizing the target speech,
and the features from all other directions are redundant.
Therefore, we add an subnet to decide how much attention should the ASR network pay to each direction.
The output of the subnet, i.e., the spatial attention $\mathbf{A}[t] = {\left(A_1[t],\ldots,A_P[t]\right)}^{\text{\sffamily T}} \in {[0,1]}^P$,
is used as the weight for the weighted average pooling across directions on the multi-direction features:
\begin{equation}
  \bar{Z}_l[t] = \sum_{p=1}^{P} {A_p[t] Z_{p,l}[t]}, \quad l = 1,2,\dots,L.
  \label{eq:att}
\end{equation}

The network predicts the spatial attention with
stacked long short-term memory (LSTM) recurrent neural networks~\cite{hochreiter_long_1997} (Figure~\ref{fig:online_att}).
The hidden states are projected to $P$ directions and normalized with a softmax layer.

Depending on when we take the output attention scores, we categorize the attention into \emph{online}, \emph{offline} and \emph{offline with controlled latency}.
The \emph{online} version outputs attention scores for each direction at each frame.
Considering that the direction of the sound source may not rapidly change,
we apply averaging with a sliding window through frames to smooth the attention prediction.
Whereas, in the \emph{offline} version, the attention scores at the last frame of the sequence is applied to all frames.
If we assume the sound source is not moving within one utterance,
the whole utterance provides more information for the network to select the correct direction than just using local frames.
The drawback of the offline version is the increased latency, which would not allow us to use this method for a streaming ASR.
Therefore, we also propose a \emph{latency-controlled} version,
which uses a short segment for attention prediction
and applies the scores at a certain frame to the whole sequence, thus constraining the latency.
The motivation of this approach is that when a trigger word is used to wake a device (e.g. ``OK google'', ``Alexa'', ``Hey Portal''), the system could rely on the wake word segment to predict the attention without adding latency to the ASR.

\subsection{Back end: Acoustic Model}

After the attention-based pooling layer, the pooled features are stacked in 8 frames and subsampled by a factor of 3. 
The stacked features are then used as the input of the acoustic model,
which predicts a sequence target posteriors of context-dependent phones or graphemes from the input feature sequence.
Our approach suits arbitrary acoustic model neural networks.
In this paper, we experiment with stacked uni-directional LSTMs as the back ends.

\section{Experiments}

We compare the attention-based deep beamforming neural network for far-field ASR with a traditional signal processing-based front end as well as fCLP with other pooling methods.

\subsection{Data}

We use an in-house anonymized dataset collected through crowdsourcing in the experiments.
The crowd-sourced workers were asked to record commands for a smart assistant on mobile devices.
Example commands include calling a contact, playing music, setting up timer/alarm, or getting time/weather information.
The average duration of the utterances is around three seconds.
More details about this dataset can be found in~\cite{le_senones_2019}.

The training data are generated from these original near-field recordings in this dataset as follows:
We randomly sample two million utterances (around 2000 hours) and simulate speech in reverberant and noisy environments.
Specifically, for a given utterance,
we first randomly simulate a room with a response time (RT60) between 200ms and 900ms,
with different locations of the device microphone array, the speech source, and the noise source in the room.
The average distance between the device microphone array and the speech source is three meters.
The microphone geometry is set to match the device used to record the evaluation data: 4 microphones roughly forming a rectangle of 6$\times$7 cm.
We then randomly sample a background noise segment from an in-house collected dataset.
Next, the clean near-field speech and the noise segment are both convolved with the simulated room impulse responses (RIR),
and added together.
We control the signal-to-noise ratio (SNR) of the resulting far-field noisy speech segment to be between 0 and 25 dB.

The evaluation data is collected by playing clean speech and noise from loudspeakers at various angles and distances, and recording with Portal+ devices.
The clean speech and the noise are different segments sampled from
the same datasets used for training.
The evaluation set includes 16500 utterances (around 16 hours).

\subsection{Implementation Details}
The baseline methods includes the log-mel feature approaches,
with or without a traditional Digital Signal Processing (DSP) front end,
and the fCLP without the spatial attention.
The DSP front end baseline consists of the typical speech enhancement algorithms,
including dereverberation, beamforming, and post-filtering. 
For the baseline method with no enhancement, the log-mel features are extracted from the first channel of the raw audio data (we can assume that which channel is selected does not affect the result).
We compute 80 log-mel features with a window size of 25 ms and shift of 10 ms on the enhanced signal and use them as the input to the acoustic model.

For the fCLP-based models, the inputs are spectrograms with the same windowing as the log-mel features.
The input is optionally dereverberated using the Weighted Prediction Error (WPE) algorithm~\cite{yoshioka_generalization_2012}.
We use the block-online implementations from the NARA-WPE library~\cite{drude_nara-wpe:_2018}.
For the beamformer front end, we use $P=10$ look directions and $L=120$ spectral features.
In addition to the attention-based pooling, we include the fCLP with no pooling (concatenate features from all directions~\cite{li_acoustic_2017}), max-pooling, and average-pooling (all directions with equal attention) for comparison.

In terms of the back end, we use a 5-layer uni-directional LSTM with 1200 units per layer for all approaches.
We delay the prediction by 10 frames, so that there are more information available for the prediction.
The output of the acoustic model is the posterior probability of 8576 context dependent graphemes ~\cite{le_senones_2019}.
We use the cross entropy as the loss function.
The front and back ends are jointly trained with the adam optimizer~\cite{kingma_adam:_2015} for 20 epochs.
The learning rate is set to 0.001 and reduced by half whenever the validation loss does not decrease at the end of an epoch.  During evaluation, we use the Weighted Finite-State Transducer (WFST) decoder~\cite{mohri_weighted_2002} with a 4-gram language model.

\section{Results}

We evaluate the performance of the proposed method and visualize its the directivity and attention patterns.

\subsection{Performance Comparison}

\begin{table}[t]
  \caption{%
    Performance of the far-field ASR systems with the spatial attention compared with the baseline approaches.
    The dereverberation is optionally applied to both the training and evaluation data for the fCLP-based approaches.
  }\label{table:online}
  \vspace{2mm}
  \centering
  \begin{tabular}{lrr}
    \toprule
    \multicolumn{1}{c}{\textbf{Front-end Approach}} & \multicolumn{2}{c}{\textbf{WER (\%)}} \\
    \midrule
    No enhancement, log-mel                         &               & 14.3             \\
    DSP, log-mel                         &               & 13.8   \\
                                         & \multicolumn{2}{c}{Dereverberation} \\
    fCLP, with pooling                   & None          & Online WPE    \\
    \cmidrule(lr){1-1} \cmidrule(lr){2-3}
    \hspace{1em}None (baseline)          & 15.6          & 12.7          \\
    \hspace{1em}Max                      & 13.8          & 12.8          \\
    \hspace{1em}Average                  & 14.3          & 13.0          \\
    \hspace{1em}Attention (online)       & 13.0          & 12.4          \\
    \cmidrule(lr){1-1}
    \hspace{1em}Attention (latency 0.5s) & 12.9          & 11.7          \\
    \hspace{1em}Attention (latency 1s)   & 12.6          & 11.5          \\
    \hspace{1em}Attention (offline)      & 12.6          & 11.3          \\
    \bottomrule
  \end{tabular}
\end{table}

\begin{figure}[t]
  \begin{center}
    \centering \sf \tiny \mathversion{sfnums}
    \includegraphics[width=\linewidth]{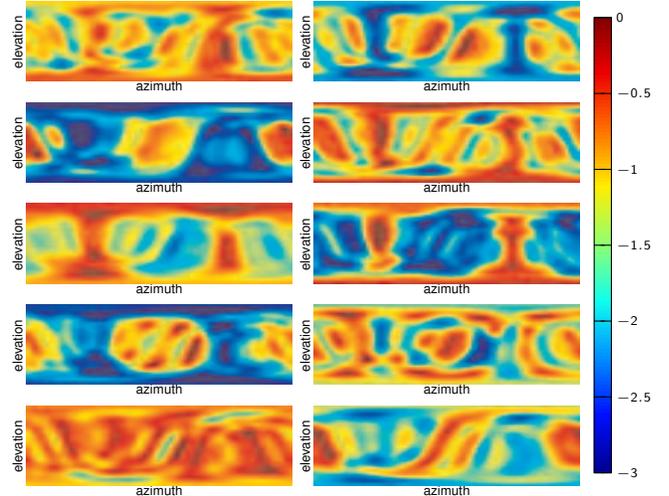}
  \end{center}
  \vspace{-8mm}
  \caption{%
    Directivity patterns of the neural beamformers trained with online attention.
    Each figure shows the directivity pattern of one spatial filter.
    The directions on the unit sphere is projected into a grid of azimuth and elevation angles.
    The color shows the gain in dB with respect to the most amplified direction (assuming the signal is white noise).
    Similar directivity patterns have been found on the offline attention (which is therefore omitted).
  }\label{fig:dir_att}
\end{figure}

\begin{figure}[t]
  \begin{center}
    \centering \sf \tiny \mathversion{sfnums}
    \includegraphics[width=\linewidth]{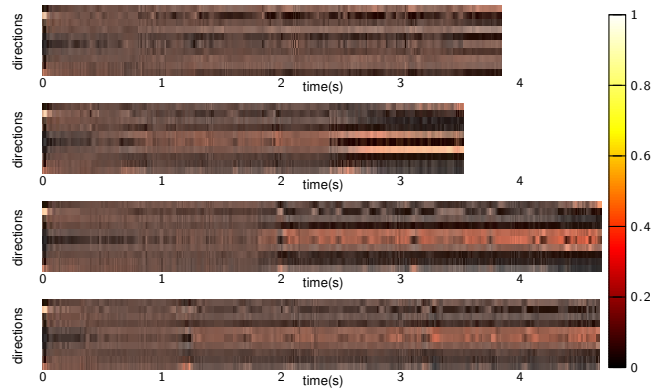}
  \end{center}
  \vspace{-8mm}
  \caption{%
    Visualization of the online spatial attention.
    It shows the attention score of each look direction (spatial filter) at each frame.
  }\label{fig:att_pattern_online}
\end{figure}

We compare different front-end approaches combined with the same back end (Table~\ref{table:online}).
The results show that the fCLP without pooling does not perform better than the log-mel approach without speech enhancement.
This may be because of the model using all the features without pooling contains more parameters, and thus could be more prone to overfit.
Adding the spatial attention let the fCLP outperform the traditional methods.
Compared to no pooling, the online attention achieves 17\% relative improvement if no dereverberation is applied.
Applying dereverberation further improves the results. However, it also reduces the gap between the online attention and no pooling.
The dereverberation reduces the potential mismatch between the training and evaluation signals,
hence mitigates the overfitting issue of the no pooling approach.

The best performance is achieved by the offline attention, which uses whole utterances to predict the attention scores.
However, the offline computation compromises the real-time responsiveness of the ASR system.
As a trade-off, the attention with controlled latency uses half second or one second to predict the attention scores,
performs almost as good as the offline attention.
Compared to no pooling, the spatial attention with one second latency achieves 9\% relative improvement.
The results support our idea of potentially using trigger word for attention prediction.

\subsection{Directivity Patterns}

We plot the directivity patterns of the neural beamformers (Figure~\ref{fig:dir_att}).
The directivity patterns show how the signal from a direction is amplified (or attenuated).
In contrary to what we expected,
the directivity patterns show that the beamformers do not amplify on a single look direction (one red blob in the figure).
Instead, each beamformer amplifies in many directions.
Moreover, the gain between the amplified and attenuated directions is small (up to around 3 dB for white noise).
This is likely due to that the number of spatial filters being much fewer than the possible signal directions.
Therefore, the neural beamformer needs to cover the unit sphere with limited filters.
Furthermore, because in our training data there are only one noise source in each utterance,
a spatial filter is effective as long as it amplifies at the speech's direction and attenuate in that noise direction,
while all other directions do not matter.

\subsection{Attention Patterns}

We plot the predicted attention scores of the online spatial attention on some of the evaluation utterances (Figure~\ref{fig:att_pattern_online}).
The patterns show that the attention of the first two seconds of an utterance is evenly distributed among all spatial filters (look directions).
It is because the network requires certain amount of information to pick the correct filters.
Unlike what we expected, even after the first two seconds, the attention is focused on a few filters instead of one.
Empirically, the averaging of features from multiple directions helps the recognition (which also can be seen by comparing the performance between average-pooling vs no pooling).
We have also found that there are some directions constantly getting more attention than the others.
Additionally, the number of directions saturates.
This is consistent with our experiments that more look directions ($P=20$) does not improve the WER.

\section{Conclusion}

We have proposed spatial attention for multi-direction deep beamforming neural networks.
The spatial attention is used to select the directions to attend to,
so that it reduces the dimension of the neural beamformer features
while keeping the most informative features of the target speech.
Our experiments with far-field ASR on a smart speaker device
show that the deep beamforming neural network with spatial attention performs up to 9\% better
than that without the attention.


\bibliographystyle{IEEEbib}
\bibliography{asr}

\end{document}